\def\year{2022}\relax
\documentclass[letterpaper]{article} 
\usepackage{aaai21}  
\usepackage{times}  
\usepackage{helvet} 
\usepackage{courier}  
\usepackage[hyphens]{url}  
\usepackage{graphicx} 
\usepackage{natbib}  
\usepackage{caption} 
\usepackage{subcaption}
\usepackage{xcolor}

\usepackage[utf8]{inputenc}
\usepackage{pbox}
\usepackage{verbatim}
\usepackage{booktabs}
\usepackage{amsmath} 
\usepackage{siunitx}
\usepackage{caption}
\usepackage{dcolumn}
\usepackage{multirow}
 \usepackage{lscape}
\usepackage[toc,page]{appendix}
\usepackage{pdflscape}

\newcommand{\astii}{^{**}}
\newcommand{\asti}{^{*}}

\title{The Hipster Paradox in Electronic Dance Music: How Musicians Trade Mainstream Success off against Alternative Status}

\author{
    Mohsen Jadidi,\textsuperscript{\rm 1,2}
    Haiko Lietz,\textsuperscript{\rm 1}
    Mattia Samory,\textsuperscript{\rm 1}
    Claudia Wagner\textsuperscript{\rm 1,3} \\
    }
    
\affiliations {

    \textsuperscript{\rm 1}GESIS -- Leibniz Institute for the Social Sciences, Cologne, Germany\\
    \textsuperscript{\rm 2}University of Koblenz-Landau, Koblenz, Germany\\
    \textsuperscript{\rm 3}RWTH Aachen University, Aachen, Germany\\
    Mohsen.Jadidi@gmail.com, \{Haiko.Lietz, Mattia.Samory, Claudia.Wagner\}@gesis.org
}
\begin{document}

\maketitle

\begin{abstract}
The hipster paradox in Electronic Dance Music is the phenomenon that commercial success is collectively considered illegitimate while serious and aspiring professional musicians strive for it. We study this behavioral dilemma using digital traces of performing live and releasing music as they are stored in the \textit{Resident Advisor}, \textit{Juno Download}, and \textit{Discogs} databases from 2001-2018. We construct network snapshots following a formal sociological approach based on bipartite networks, and we use network positions to explain success in regression models of artistic careers. We find evidence for a structural trade-off among success and autonomy. Musicians in EDM embed into exclusive performance-based communities for autonomy but, in earlier career stages, seek the mainstream for commercial success. Our approach highlights how Computational Social Science can benefit from a close connection of data analysis and theory.

\end{abstract}

\section*{Introduction}

Counter-cultural and anti-establishment fields legitimize themselves by distancing from the mainstream. Yet, to sustain their careers and achieve economic success, cultural producers in such fields need to strive for widespread recognition for their work. Approaching mainstream success while not becoming mainstream themselves, running the risk of alienating supporters from their subculture and being labeled as a ``sell out,'' is the paradox that subcultural producers face.
We refer to it as the \emph{hipster paradox}, borrowing the term from the phenomenon that the hipster subculture blends mainstream and alternative lifestyles \citep{greif_positions_2010}.

Scholars in sociology and the \emph{science of success} have studied what makes for a successful career among Jazz \citep{pinheiro2009all} and Punk musicians \citep{10.2307/j.ctt18mvkkz}, painters \citep{giuffre1999sandpiles}, and writers
\citep{DENOOY2003305}. In recent years, a magnitude of studies have investigated the working condition and success of creative careers using large-scale digital behavioral data
\citep{rossman2010d, allington2015networks, janosov2020elites}.
Yet, we know little about how producers in counter-cultures deal with the dilemma the hipster paradox poses to agents in creative industries.

Electronic Dance Music (EDM) makes for an interesting case study because its history is one of non-conformity with mainstream music culture: mainly white Rock music. It started as a collective action by those who felt alienated by the mainstream: mainly the black and gay population \citep{mcleod2001genres}. EDM's increasing popularity in the last decade has brought it from the margin and underground culture to an industry with a global value of 7.3 billion US dollars in 2019 \citep{watson_ims_2020}.
Yet, autonomy from mainstream values is itself a central value in EDM. Unregulated, unlicensed, anti-establishment, and exclusive parties, organized by communities of enthusiast, served as a safe space for personal expression and liberty
\citep{anderson_rave_2007}.

\begin{figure*}[t]
    \begin{subfigure}{0.24\textwidth}
        \includegraphics[width=\textwidth]{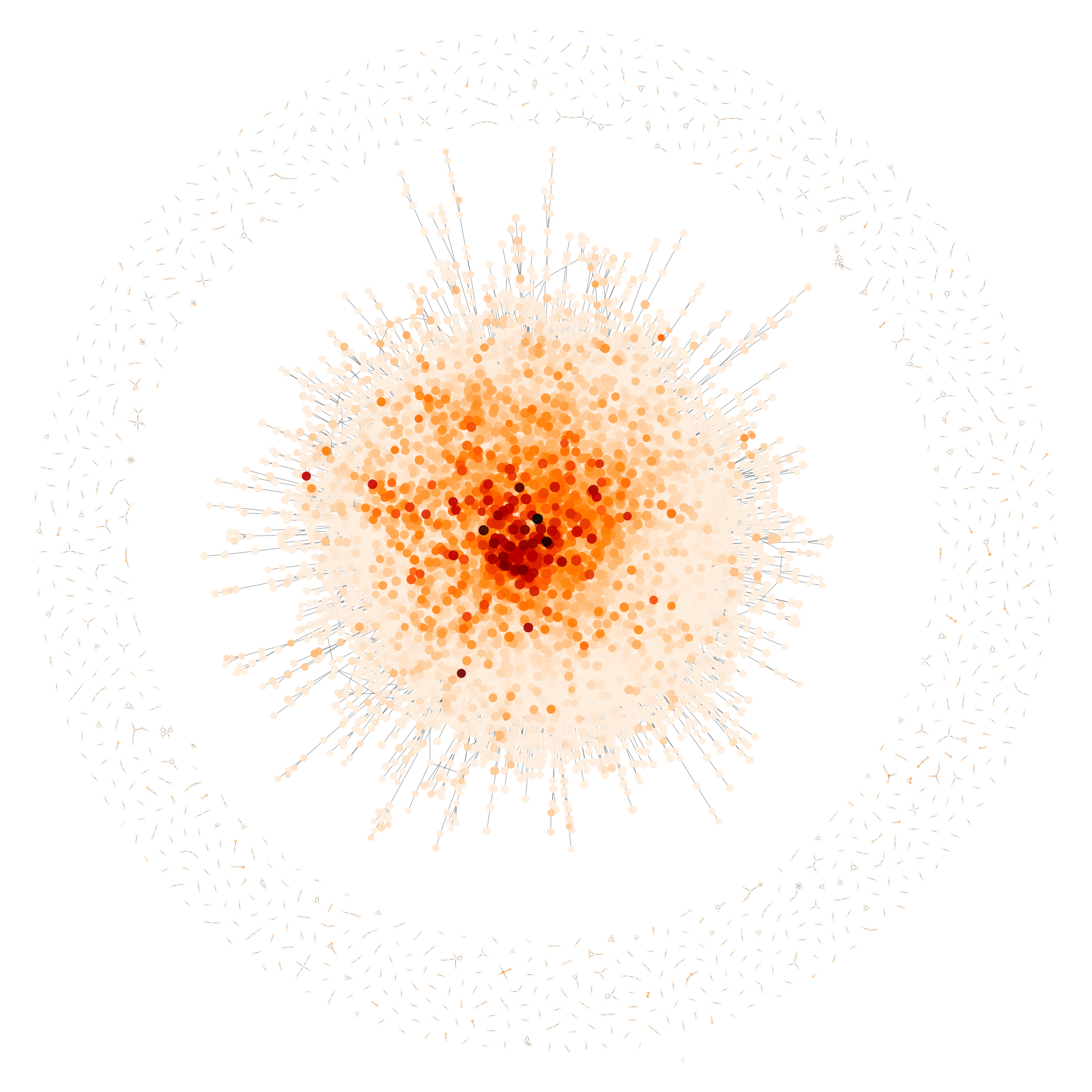}
        \caption{Co-gig}
    \end{subfigure}
    \begin{subfigure}{0.24\textwidth}
        \includegraphics[width=\textwidth]{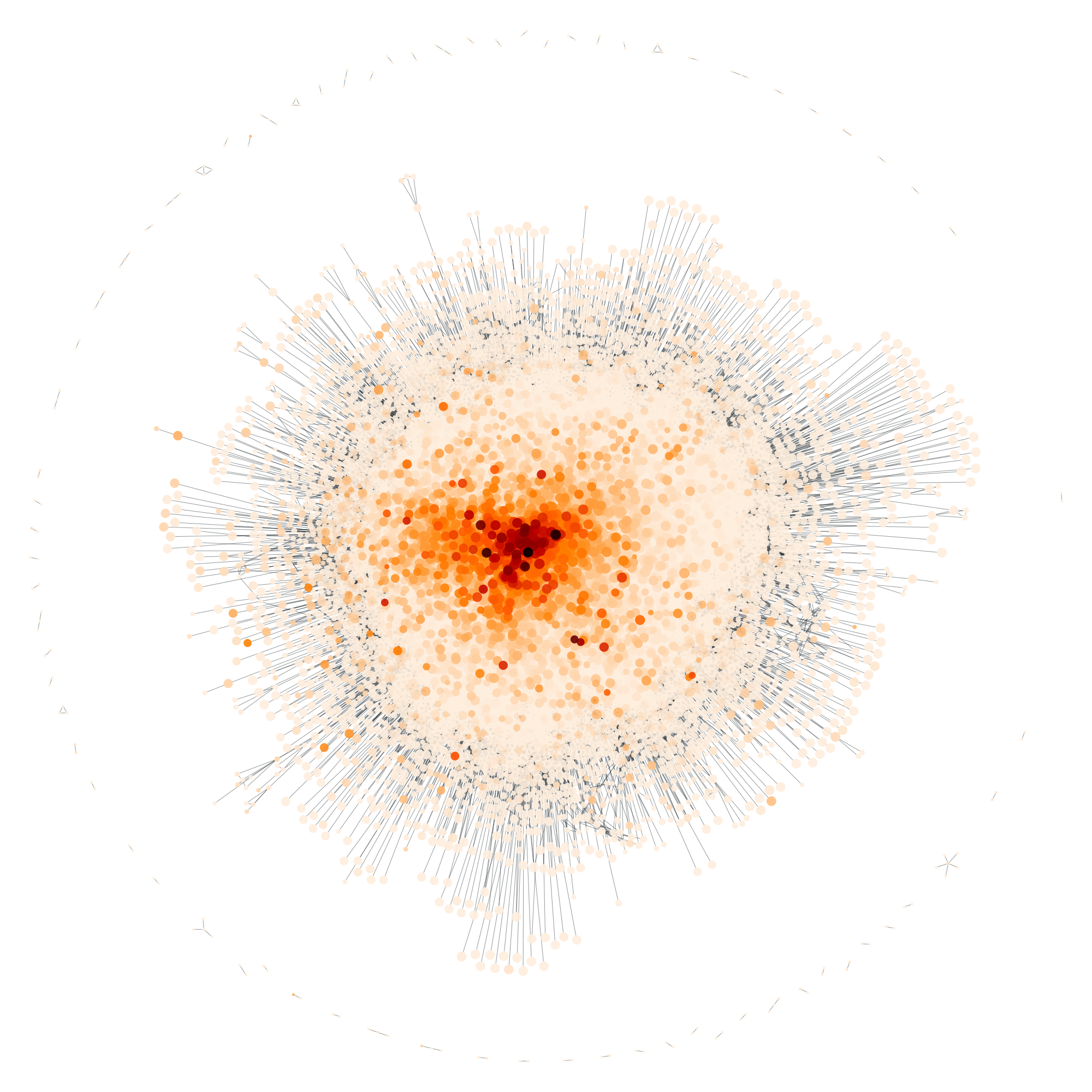}
        \caption{Co-venue}
    \end{subfigure}
    \begin{subfigure}{0.24\textwidth}
        \includegraphics[width=\textwidth]{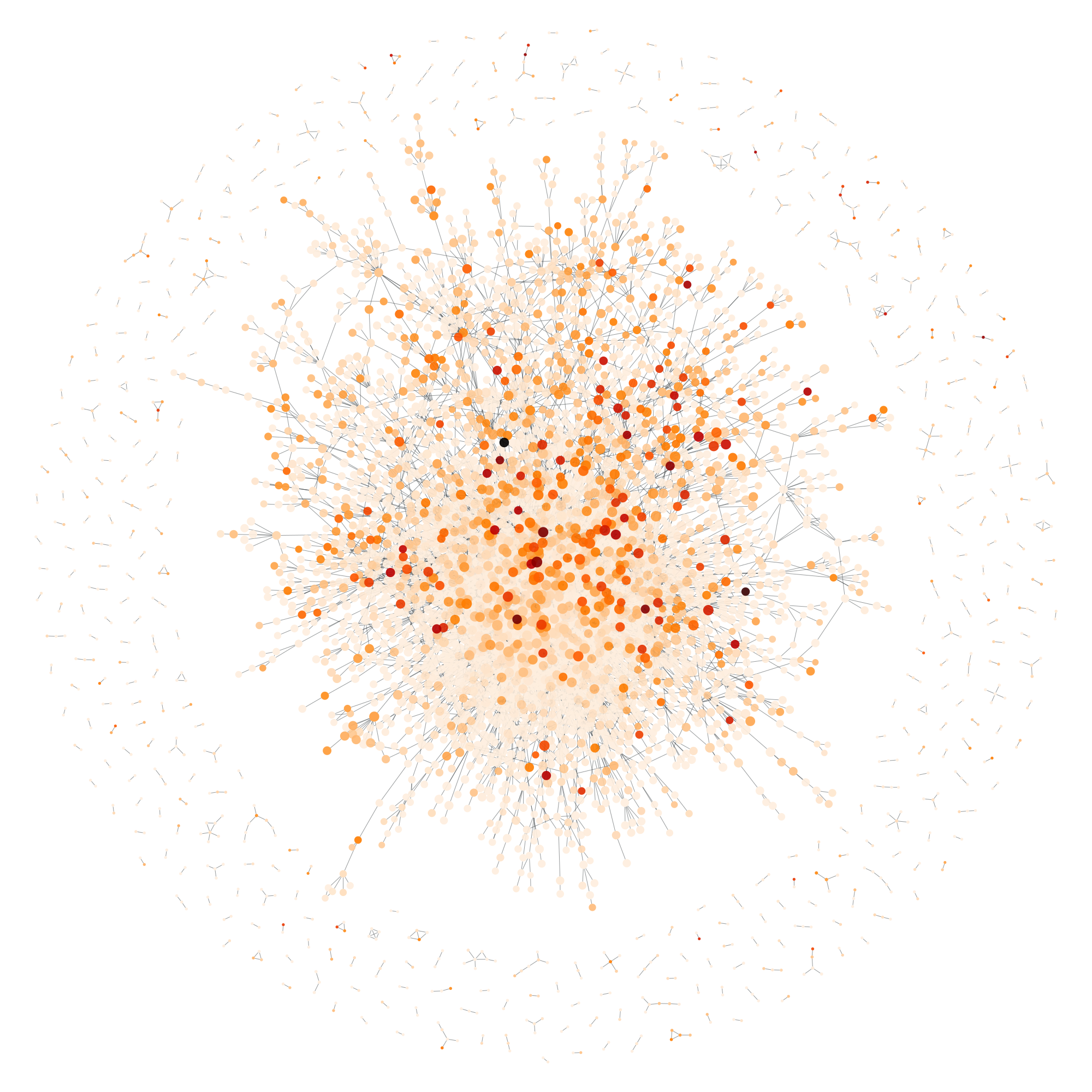}
        \caption{Co-label}
    \end{subfigure}
    \begin{subfigure}{0.24\textwidth}
        \includegraphics[width=\textwidth]{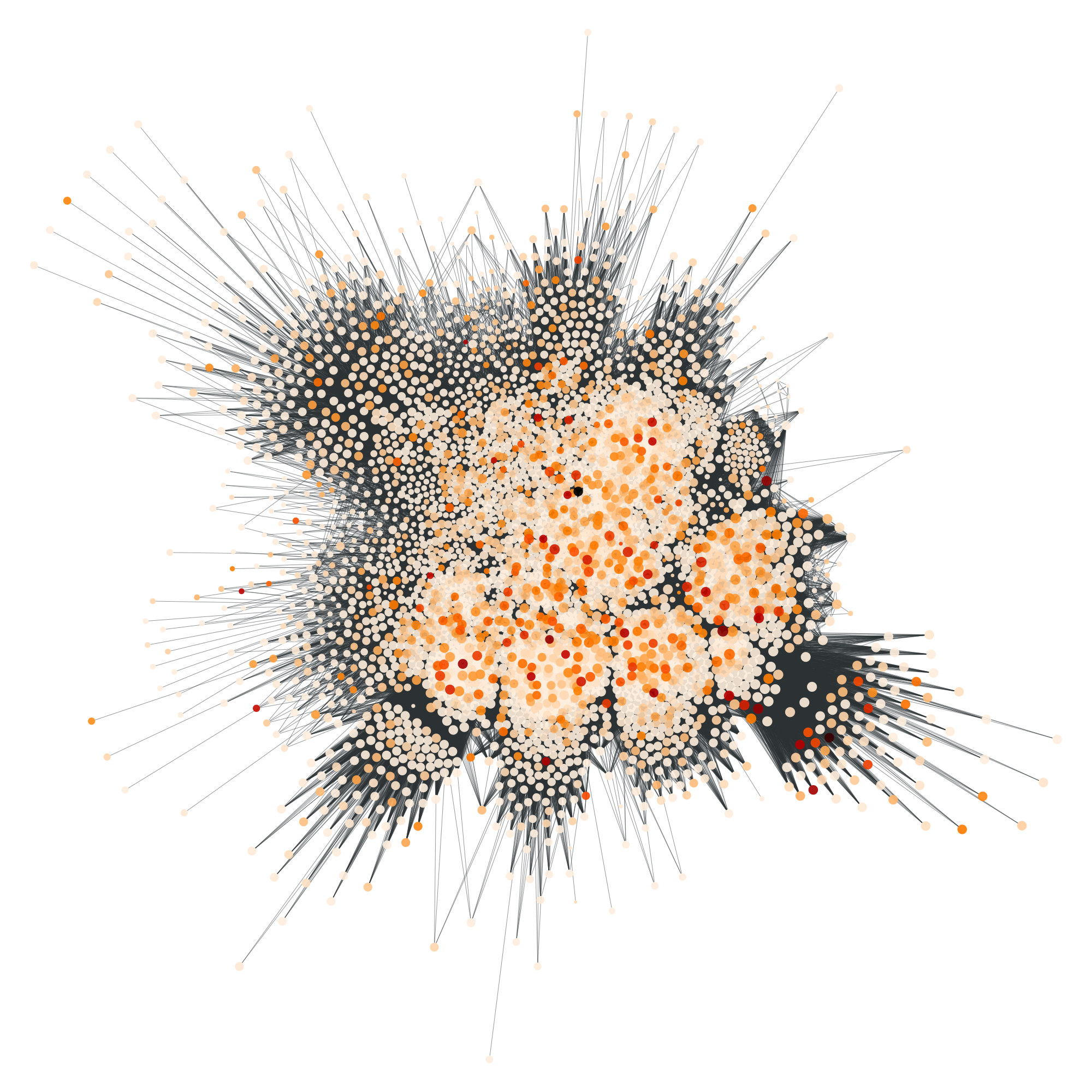}
        \caption{Co-style}
    \end{subfigure}
    \caption{
    Networks of musicians connected by (a) co-performing at gigs, (b) co-performing in clubs and other locations, (c) co-releasing on music labels, and (d) co-releasing in music styles. Networks are largest connected components with insignificant ties and isolated nodes removed. Node size is proportional to how close a musician is to all others (closeness centrality). Node color gives a musician's success in terms of the distance traveled between live performances (the darker the more successful). Intuitively, musicians of international renown are in demand in venues that are distant from each other. These snapshots uncover that successful musicians follow the mainstream by taking central positions in networks built on gigs and venues. In this paper, we show that this strategy is associated with success in early career stages. Snapshots are for the 2013-2015 period.
    }
    \label{fig:networks}
\end{figure*}

Large-scale behavioral data from digital platforms enable unobtrusive, longitudinal analysis that may help to uncover behavioral patterns and mechanisms. Such studies on EDM have highlighted the importance of community embeddedness for value creation and success \citep{allington2015networks, janosov2020elites}.
However, insights into how musicians deal with the hipster paradox are mainly derived from qualitative interviews with musicians. These diagnose a success/autonomy trade-off that consists of rooting commercial practices in exclusive and alternative performance-based communities \citep{reitsamer2011diy, lange_value_2013, wreo20632, wilderom_intersecting_2019}.
The ethnographic method allows for in-depth insights, but it relies on retrospective accounts of field participants that suffer from memory and desirability biases.


In this paper, we study the hipster paradox in EDM using large-scale and longitudinal digital traces of musicians. Grounding our observations in the careers of over 4,000 artists over almost two decades, we study how their relationship with mainstream appeal affects their success. Using digital trace data has the benefit that our observations are unobtrusive accounts that unfold over time.
Inspired by sociological field theory \citep{bourdieu_field_1993}, we identify two primary practices that embed artists within the EDM subculture and are not mainstream or alternative \emph{per se}: performing live and releasing music. 
Whereas mainstream labels and live venues are a conduit to widespread popularity and economic success, alternative releases and performances reinforce and legitimize the artists' belonging to the EDM subculture.

How important is it for musicians to be embedded into a community? How important is it to belong to the mainstream? Are bridging or redundancy-avoiding strategies associated with success? And how does all that change over an artist's career?
To answer these questions, we construct a large data corpus by harvesting the \textit{Resident Advisor}, \textit{Juno Download}, and \textit{Discogs} platforms. For the 2001-2015 period of observation, we construct four analytical networks that convey how similar musicians are in terms of practicing EDM \citep{batagelj_bibliographic_2013, edelmann_formal_2018}. We quantify positions in these networks, devise a measure of success that is based on long-distance travels, and regress success on network variables in linear mixed models.
Figure \ref{fig:networks} gives an impression of these networks and the position of successful musicians.

We find evidence of a structural trade-off between revenue and autonomy. Musicians in EDM embed into exclusive performance-based communities for autonomy but, in earlier career stages, seek the mainstream for commercial success.
Our results show that successful musicians gain a sufficient support base early in their careers at the risk of ``selling out,'' while established artists that assert their alternative status find long-term success.

\section*{Related Work}

\subsection*{Electronic Dance Music}

The hipster paradox can be rooted in the sociological theory of ``fields of cultural production,'' a framing that is useful for understanding the conflict of art and money. According to this idea, legitimacy in fields of art (i.e., subfield of restricted production) springs from autonomy from the economic order (i.e., from subfield of large-scale production) \citep[][ch.~1]{bourdieu_field_1993}. 
In EDM, the relationship of art and money is complex (and subject to our modeling). The history of EDM shows that a polar distinction between those that do ``art for art's sake'' and those that work for the ``creative industry'' are too simple. For example, the EDM subfield in the UK is much more centralized and commercialized than the US subfield, but it emerged from the latter's reluctance to partner up with the record industry \citep{wilderom_intersecting_2019}.

Nowadays, EDM is home to the ``notion that, equipped with the right set of tools, skills, and talent, one individual can `make it' alone''  \citep[][p.~152]{wreo20632}.
\citet{reitsamer2011diy} finds that musicians in EDM seem to embody this ``Me Inc.'' ideology, that is, they do strive for commercial success day by day, and concludes that it calls into question the supposed autonomy of cultural producers.
This situation makes the hipster paradox an existential problem for musicians. 

There are two main practices in EDM that allow them to face the dilemma. The practice of \emph{performing live} is strongly related to the notion that EDM enshrines a love of music and dancing. In \emph{gigs} such as club nights and raves, performance and participation meld, and music acts as a gravitational force for social relations \citep{turino2009four}. 

Serious and aspiring professional musicians must carefully choose in which \emph{venues} to play. On the one hand, larger venues pay higher wages, but, on the other hand, since mass production is considered ``selling out,'' performances in big clubs are endowed with a negative label \citep[][ch.~4]{wreo20632}.

Interviews with musicians suggest that they address the paradox via a particular kind of network sociality: ``As individualistic entrepreneurs, grassroots musicians often find themselves in weak positions, having less power to negotiate conflicts, bargain for better opportunities, and navigate the social structures and groups that organise EDM musical activities. To compensate, many aspiring professional participants join networks who function as `defensive exclusionary networks' ..., and in the process distance themselves from others.'' \citep[][p.~156]{wreo20632}
During live performance events, strategic relationships occur in settings that correspond to musicians' natural state of being \citep{lange_value_2013}.
Musicians embed into systems of intersubjective ties that are ``informational, ephemeral but intense, and ...  characterized by an assimilation of work and play.'' \citep[][p.~71]{wittel2001toward}. Since these networks maintain familiarity and mutual valuation, commercial success is not stigmatized \citep{reitsamer2011diy}.
According to field theory, these are the ``natural'' environments that allow artists to be authentic and escape the hipster paradox \citep{michael_its_2015}.
Local club scenes are the vivid faces of these dynamics. For one, they form around geographical locations where cities like London and Berlin take core positions in the field \citep{allington2015networks}.

Besides performing live, \emph{releasing music} is the other main practice in EDM.
Songs and records are a way to express autonomy. Other than performing, which requires at least access to a venue, musicians are, in principle, free to produce in whatever \emph{style} or music genre they want. Musicians are free to just release their music online or to start their own label \citep{reitsamer2011diy}. This informal ``do-it-yourself'' culture drives the evolutionary dynamics of EDM. For example, ``drum-n-bass'' is a main genre that differentiated into ``abstract drum-n-bass,'' ``ambient drum-n-bass,'' and ``intelligent drum-n-bass'' \citep[][p.~60]{mcleod2001genres}.
Like venues, styles are crystal nuclei of exclusionary practices in communities \citep[][p.~219]{wreo20632}.

By released music, musicians demonstrate their seriousness and gain access to the inner social circles of communities which opens new pathways to making a career \citep{mcleod2001genres, reitsamer2011diy}.
To produce and release at a large scale, musicians have to secure deals with music \emph{labels}. Labels function as gate-keepers of the creative industry: They sift through the pool of cultural producers and select those that are promising to meet the current taste of the community or field. This asymmetric power over the boundaries gives them influence over the tastes, opinions, and reputations of producers, performers, and participants \citep{kennedy2008getting, reitsamer2011diy}.
It has been found that a small fraction of star artists help other musicians into top ranks via mentorship and recording collaboration. Which musicians these are is, in turn, influenced by their styles, that means, changes in the social cores of communities mirror the cultural drift of styles \citep{janosov2020elites}.

The literature has identified ways in which EDM musicians employ performance-based practices to navigate the dilemma. We build upon this literature, finding corroborating evidence of how embedding in communities of musicians facilitates this process. We expand upon these insights, inquire about the role of practices related to releasing music, and show how stages in EDM careers mediate which of these practices are successful.
Studies on the hipster paradox in EDM highlight the importance of network effects. In contrast to these ethnographic studies, we perform an empirical analysis of the network of musicians based on the digital traces of their practices. Hence, we next discuss the related literature at the intersection of network science, art, and success.

\subsection*{Network Analysis of Fields of Art}

Networks have been shown to be apt representations of fields. Most abstractly, a field is a space of relations among positions. Fields govern individuals' practices, and they manifest as social networks. The power of graph-theoretical approaches is that they make positions amenable to measurement and computation \citep{DENOOY2003305, bottero2011worlds}.
One way to construct these analytical structures is by way of bipartite (two-mode) networks. By modeling practices as relationships of agents and symbolic facts (e.g., music venues or styles), formal frameworks allow for constructing fields as networks from practices \citep{edelmann_formal_2018,blasius2020empirical}.
This approach involves a projection of the two-mode network to a binary or weighted one-mode network \citep{batagelj_bibliographic_2013}. Agents with similar patterns of choices in the initial two-mode network have similar patterns of ties and, hence, similar positions in the projected one-mode network. The structure of this network can then be analyzed and visualized using the graph-theoretical repertoire of Social Network Science \citep{ moody_structural_2003,borgatti_graphtheoretic_2006}. 

Music involves a series of relations between a variety of agents such as artists, mentors, recording studios, labels, distributing companies, promoters, music venues, audiences, and critics \citep{small1999musicking}.
A number of studies uses networks to explore music fields and musicians’ careers \citep{allington2015networks,Crossley_2020,rachel_metal,millward_britpop}.

For example, an analysis of the bipartite network of artists and festivals shows that Turkey’s Metal music field exhibits a core-periphery structure. Bands with a stronger affiliation to the Rock style, a larger number of festivals played, and support from major labels are more likely to occupy central positions in the network \citep{rachel_metal}. Similar work on Punk \citep{10.2307/j.ctt18mvkkz} and Jazz \citep{vedres2017forbidden} suggests that artists who occupy central positions in co-gig networks and form open cliques have higher chances of success.
However, most studies consider only one aspect of musicians' careers, namely the affiliation to either gigs, venues, labels, or music styles. Our work contributes to this line of research by analyzing the career of EDM artists using all four networks.

\section*{Materials and Methods}

\subsection*{Datasets}
\begin{table*}[t]
    \centering
    \begin{tabular}{lrrrrrr}
     & \textbf{Musicians} & \textbf{Gigs} & \textbf{Venues} & \textbf{Releases} & \textbf{Labels} & \textbf{Styles} \\ \hline
    \textbf{Live performances} &  &  &  &  &  &  \\
    ~~\textit{Resident Advisor} (RA) & 63,543 & 728,850 & 50,410 & - & - & - \\
    \textbf{Releases} & 39,042  &  &  &   &  &  \\
    ~~\textit{Juno Download} (JD) & 35,844  & - & - & 259,147 & 30,488 & 69 \\
    ~~\textit{Discogs} & 23,663  & - & - & 160,130  & 30,281 & 339 \\
    \textbf{Total} & 63,543 & 728,850 & 50,410 & 332,162 & 39,661 & 347 \\ \hline \hline
    \end{tabular}
    \caption{Dataset statistics. Each dataset offers partial information about practices of artists in EDM. RA consist of a larger number of artists and serves as the primary source for musicians. JD and \textit{Discogs} together provide release information for about half of the musicians. While JD has more releases, \textit{Discogs} provides richer information on music styles.}
    \label{tab:data_stats}
\end{table*}

Our research design calls for measuring the field of EDM via the practices that constitute it.

To study the hipster paradox in a large-scale quantitative way, we collect data from digital platforms. 

While all datasets come with limitations, which we discuss towards the end of the paper, these are especially suitable to our design because they capture the digital traces left by the practices of performing live and releasing music.

We build a corpus of traces from three platforms, each providing partial information (\textit{Resident Advisor} for performances, \textit{Juno Download} and \textit{Discogs} for releases). Once combined, this corpus offers a holistic view of the field.

\subsubsection*{Live Performances}

We use \textit{Resident Advisor} (RA, \url{residentadvisor.net}) as a primary source for selecting a large sample population of EDM musicians and information about their live performances. RA is an online music magazine and platform dedicated to EDM. It serves as one of the main information hubs for EDM events and culture worldwide.

Musician profiles contain information about their ``gigography'' including event venue, date, and lineups. Similarly, each venue has a profile page that includes information such as its address, social media links, and archived past events.
We infer the geo-coordinates and locations of venues by using the combination of four geo-location APIs, namely the Nominatim, HERE, Google, and GeoNames APIs. For evaluation purposes, we manually assigned the city to 150 venue and found $78\%$ correct assignment from the APIs.

\subsubsection*{Music Releases}

We compile a discography of musicians by combining data from two major online music  discographies and stores.
\textit{Juno Download} (JD, \url{junodownload.com}) is considered one of the largest independent dance music download stores worldwide. It provides a large catalog of electronic music styles with over 6 million tracks. Each track is attributed with artist name, label name, release name, release date, and music genre(s).
\textit{Discogs} (\url{discogs.com}) is a crowdsourced discography platform, the largest and most comprehensive music database and marketplace with 10 million releases across various genres. With a share of 14.26\%, electronic music is the second largest genre (after Rock with 23.68\%) in the platform.\footnote{Matt Larner, ``State of Discogs 2017,'' \textit{Discogs BLOG}, February 14, 2018, \url{https://blog.discogs.com/en/state-of-discogs-2017/}, retrieved June 11, 2021.} The platform provides information about musicians and bands, namely a short biography, social media and internet pages (e.g., Wikipedia, personal website), band members, aliases, and name variations.

To identify RA artists in JD and collect their discographies, we query the website using artist names extracted from RA.

To find the \textit{Discogs} page for each RA artist, we first check if there is a link in the artist's RA profile page.  For the remaining artists, we use the \textit{Discogs} search API to query for artist names. For musicians with multiple aliases or projects, we combine all the releases under the name with the highest number of gigs.
To avoid name ambiguities and duplicate entries, we match label names from JD and \textit{Discogs} using release and artist names.

\subsubsection*{Final Sample and Period of Observation}

Up until 2000, the number of active musicians in RA is smaller than 1,000. All data was collected in 2018. Hence, we limit our period of observation to 2001-2018. Table \ref{tab:data_stats} reports the numbers of musicians, gigs, venues, releases, labels, and styles derived from these practices. The full sample consists of 63,543 musicians that play 11 gigs on average. For 61\% of them, we also found releases.
Those musicians that produce music release 9 pieces (either single songs or full records) on average. \footnote{Data access: \url{ https://doi.org/10.7802/2360}}

\subsection*{Methods}
\begin{figure*}[t]
  \centering
  \includegraphics[width=1\textwidth]{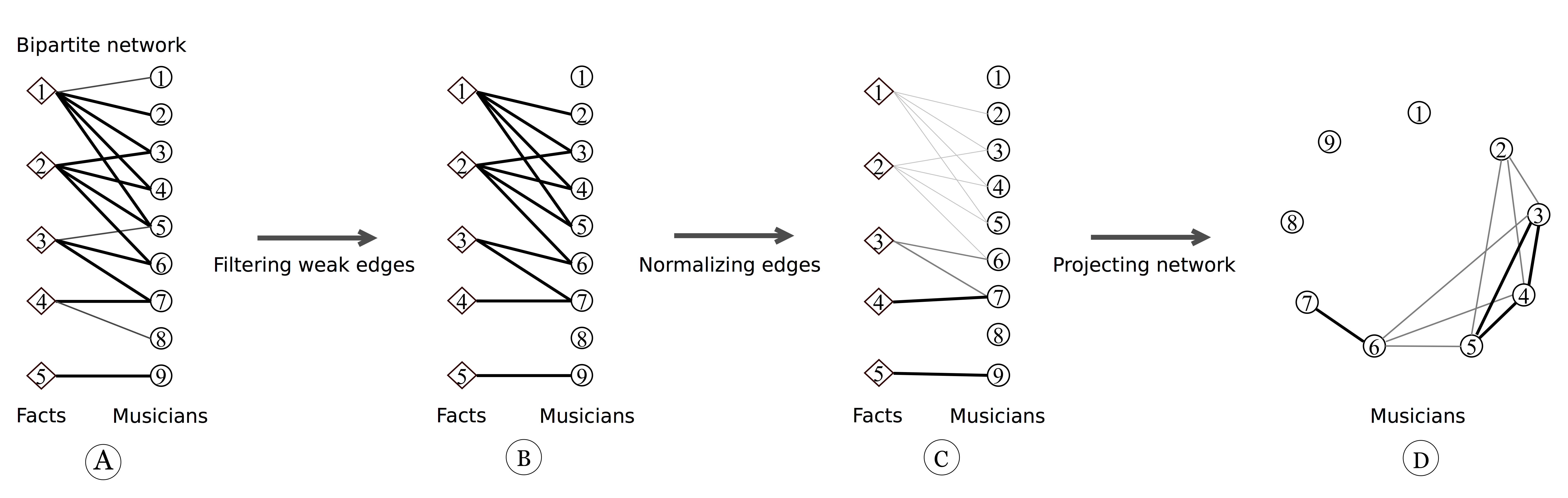}
  \caption{Construction of networks representing the field of EDM from bipartite networks representing practices in EDM. (A) Bipartite networks of facts (EDM venues, labels, or styles) and musicians with weighted edges (i.e., facts can be selected multiple times). (A→B) Weak edges are removed where a fact is selected only once. (B→C) Edges are normalized so all facts have unit weighted degree. (C→D) The network is projected to obtain the network of musicians where edge weights give their similarity in terms of selecting the same facts (performing in the same venues, releasing on the same label, or releasing in the same style). Thicker and darker edges indicate larger weights.}
  \label{fig:network_schema}
\end{figure*}

\subsubsection*{Network Construction}

To analyze fields as networks, we follow a formal approach \citep{edelmann_formal_2018}. It consists of representing the traces of practices in bipartite networks where one part is a \emph{musician}, an artist who performs live and may also release music. The other part is a \emph{fact} (gig, venue, label, or style).
Bipartite networks of musicians and gigs or venues derive from the performing practice; bipartite networks of musicians and labels or styles derive from the releasing practice.

Analytical networks representative of the field of EDM are constructed from these bipartite networks \citep{batagelj_bibliographic_2013}. Figure \ref{fig:network_schema} schematizes their construction.
There are two preprocessing steps.
First, we remove all facts that musicians chose only once since those introduce noise.
Second, we normalize edges in bipartite networks in such a way that the number of selections by all musicians sums to one for all facts.
The analytical networks are then obtained by projecting bipartite networks in such a way that musicians become the nodes.
For the creation of gigs networks, the first preprocessing step is different because the original edges are not weighted (a gig is a one-time event). Instead, we remove gigs with only one musician. A check reveals that such events are mainly data artifacts caused by missing information in the lineup listings. We also remove gigs (such as festivals) with a large number of musicians (three standard deviations over the mean). These are rare events that entail many but, due to normalization, weak edges that overshadow network analysis.
This method results in musician \emph{co-gig}, \emph{co-venue}, \emph{co-label}, and \emph{co-style} networks (snapshots for the 2013-2015 window are depicted in figure \ref{fig:networks}).

As a result of normalization, two musicians can be similar either if they co-perform in many popular gigs or venues, or if they co-perform in fewer but more alternative ones (and similarly so for releasing music on labels and in styles).
Correspondingly, communities can emerge in two different ways that map to mainstream and alternative practices.
We shall give an example. In figure \ref{fig:network_schema}D, there are two communities constituted by strong ties: The first consists of nodes 3, 4, and 5; the second consists of nodes 6 and 7. The first is a \emph{mainstream community} since it derives from all nodes selecting the popular facts 1 and 2; while the second is an \emph{alternative community} that derives from the selective focus on the otherwise unpopular fact 3.

\subsubsection*{Cohorts and Careers}

Following a longitudinal research design means that we construct the four networks described above for sliding time windows of three years.
A \emph{career} is then a sequence of positions in these networks \citep[p.~18]{bourdieu_field_1993}.
Our period of observation is 2001-2018. Musicians are the units of observation. Musician $a$ belongs to a \emph{cohort} defined by the first year $t_a$ in which they perform live. In year $t\geq t_a$, a musician has a \emph{career age} $\tau(t)=t-t_a$. We differentiate among different \emph{career stages}. A musician can be in the early stage ($\tau<5$), mid stage ($5\leq\tau<10$), or late stage ($10\leq\tau$).

We also attribute musicians to one of five success-based \emph{career types}: stable successful, stable mediocre, stable unsuccessful, upward, and downward. To do so, we use travel distance (which we introduce in section ``Measuring Success'') to compute the percentile rank $r_a(t)\in\{1, 2, 3, 4, 5\}$ of each musician in a given year (musicians in the first 20\% quantile have rank 1, in the second 20\% quantile rank 2, ...). Next, we compute the average rank $\overline{r}_a$ over a career and the rank difference $\delta_a$ between the first and last years of a career. We consider careers with  $\overline{r} \leq 2$ as ``stable successful,'' $\overline{r} \geq 4$ as ``stable unsuccessful,'' $\delta > 0$ as ``upward,'' $\delta < 0$ ``downward,'' and the remaining as ``stable mediocre.'' Upward and downward careers do not include careers already assigned to one of the first two categories.
As Figure \ref{fig:career_types} shows, ``stable successful'' is the largest category. 

\subsection*{Research Design}
\subsubsection*{Research Questions}

The literature on EDM proposes that musicians solve the dilemma posed by the hipster paradox by embedding into alternative and exclusive communities in which they can pursue commercial activities without being stigmatized \citep{reitsamer2011diy, wreo20632}.
Quite generally, \emph{communities} are cohesive social formations that have the purpose of reducing uncertainties for its constituents \citep{white_identity_2008}.
In EDM, musicians join communities for informal reasons (i.e., the love of music and dancing) and formal reasons (e.g., to strategically forge ties to market intermediaries) \citep{turino2009four, reynolds2013energy, janosov2020elites}.
First, we want to know if there is empirical evidence for the cohesive nature of EDM.

\textbf{Research Question 1:} To what extent is community embeddedness associated with success?

Next, we address the aspect of belonging to the \emph{mainstream} culture. As we have seen in the methods section, communities can have their origins in both mainstream and alternative practices. Is it true that only the alternative path leads to success, as the literature suggests? Or is the mainstream path also viable, despite its inherent risk of losing legitimacy?

\textbf{Research Question 2:} To what extent is mainstream belonging associated with success?

An important part of the explanation how musicians solve the dilemma they face is that the communities they embed into emerge from exclusionary practices, that is, musicians distance themselves from others \citep[][p.~152]{wreo20632}.
This implies that successful musicians take positions in communities that have rather impermeable boundaries. Embedding into multiple communities would then not be associated with success. On the other hand, positions in boundaries can be sources of creativity and success thanks to the opportunities of \emph{bridging} structural holes that exist between communities \citep{burt_structural_1992}.

\textbf{Research Question 3:} To what extent is bridging associated with success?

The structure of a musician's immediate network neighborhood may also have an effect. It has been shown that dense ego networks are detrimental to creativity, likely because they are correlates of rather indistinguishable and redundant node neighborhoods \citep{uzzi2005collaboration}. As such, they have a \emph{constraining} effect on a node \citep{latora_social_2013}.

\textbf{Research Question 4:} To what extent is constraint associated with success?

Finally, the literature, being largely based on interviews and ethnographic work, has not touched upon how changes in network positions may be associated with success over musicians' careers. 
That means, we seek answers to the questions above by differentiating between the early, mid, and late career stages of musicians.

\subsubsection*{Measuring Success}

We derive the success measure from live performances. Live performance is the main source of income in popular music \citep{montoro2011live} and particularly in EDM \citep[][ch.~4]{wreo20632}. Our rationale is that musicians who perform in gigs around the world cover long geographical distances.
Our measure is based on the trajectory of musicians' travels among gig locations. Each venue has a dedicated page in RA. We use its address to obtain the city where it is located. Let $C_a=\{c_1, c_2,...,c_N\}$ be the \emph{travel trajectory} of musician $a$ who makes $N$ visits to cities $c$ ordered in time. The same city can be visited multiple times. The success variable is then the summed travel distance $d_a=\sum_{i=1}^{N-1}\epsilon(c_i,c_{i+1})$ where $\epsilon$ is the Euclidean distance function. 

This proxy for success finds anecdotal validation in the fact that the most-traveled musicians are indeed enormously successful acts, and top EDM musicians Tiësto and Paul van Dyk lead the ranking even before Rock icons Bob Dylan and Metallica.\footnote{Jacob Shamsian, ``The 10 most-traveled musicians have toured over 11 million miles around the world — here's the full list,'' \textit{Insider}, February 15, 2017, \url{https://www.insider.com/musicians-who-travel-the-most-2017-2}, retrieved September 13, 2021.}
The variable also passes a formal evaluation test: It is able to predict which musician belongs to the top 100 in two annual international ranking polls. The average predictive accuracy is $85\%$ from 2008-2018, on average.
A comparison with other travel-based success measures shows that it is important to consider the order of city visits. This is mirrored in reality where it is common practice by grassroots artists who build their music career next to a day job to arrange for multiple live performances when they travel to far-away cities. This way, they can reach larger audiences and save time and money.

\subsubsection*{Measuring Positions in the Field}

The advantage of our network approach to field theory is that we can operationalize the four different types of positions addressed by research questions 1-4. The first two types serve to diagnose the importance of \emph{network closure} for success.
The core positions in a network represent its mainstream behavior.
We operationalize the construct of mainstream belonging as the closeness centrality in a network. The closeness of a node is the inverted sum of its distances to all other nodes \citep{opsahl_node_2010}. It is close to 1 for core nodes and close to 0 for peripheral nodes.
This is a global measure because a node's position is characterized with respect to to all other nodes.

Communities are cohesive network substructures with the density (i.e., the ratio of the numbers of observed and possible ties) increasing from the periphery of a community to its core \citep{moody_structural_2003}. We operationalize the construct of community embeddedness as the maximum $k$-core that a node belongs to, where the $k$-core is a maximal subgraph whose nodes are all connected to at least $k$ others \citep{batagelj_fast_2011}. Musicians in the core (periphery) of a community will have large (small) values.
Compared to the global closeness centrality measure, this is a local measure because it takes nodes at an intermediate distance of an observed node into account.
While closeness centrality makes use of edge weights, the $k$-core algorithm assumes an unweighted graph.

The other two types of positions refer to the importance of \emph{network openness} for success. The first is bridging which we operationalize as node betweenness centrality, the extent to which the shortest paths among all node pairs pass through a node \citep{brandes_centrality_2007}. 
Again, we contrast this global measure with a local one: The clustering coefficient \citep{watts_collective_1998} is our measure for the last construct of constraint. It is close to 1 (0) for strongly (weakly) constrained nodes. Note that this is the only network variable where an inversely proportional relationship with success is expected.
Both measures, bridging and constraint, are computed using edge weights.

\subsubsection*{Linear Regression of Success}

In probing the trade-off among success and autonomy, we treat the two constructs differently. Given that success is the outcome of collective field dynamics, and is well-defined we can directly measure it. Hence, we use success as the dependent variable and regress it on 16 independent network variables (4 types of positions for 4 analytical networks) and baseline variables. Autonomy, on the other hand, is the practice of separation from the central norms and cultural power. Here, autonomy is then captured via combinations of network variables that reflect the adoption of such practices. The analysis is longitudinal, that is, we use independent variables aggregated in 3-year time windows to explain success in the ensuing 3 years: The independent variables are computed for rolling time windows $[t-2, t]$; The dependent success variable is computed for travel trajectories in windows $[t+1, t+3]$. Observations are collected for $t\in\{2003, 2004, ..., 2015\}$. We exclude musicians that never reach a career age of 5 years as well as musicians for which there are less than 5 observations. This way, we put a focus on serious and aspiring professional musicians \citep[][p.~2]{wreo20632} and provide reasonable numbers of observations to capture trends and variations in careers.
To mitigate the impact of censoring bias in our analysis, we exclude musicians with at least one gig or release before 2001.
\begin{figure*}
    \minipage{0.2\textwidth}
        \includegraphics[width=\linewidth]{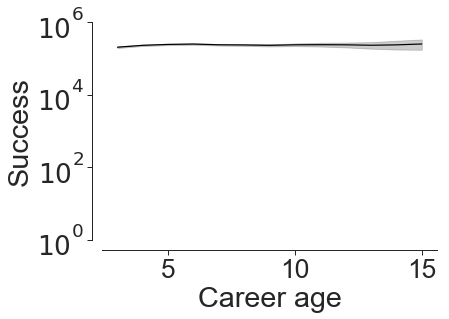}
        \subcaption{Stable successful}\label{fig:career_cat1}
    \endminipage\hfill
    \minipage{0.2\textwidth}%
        \includegraphics[width=\linewidth]{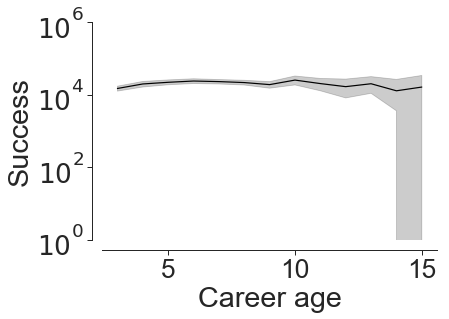}
        \subcaption{Stable mediocre}\label{fig:career_cat5}
    \endminipage\hfill
    \minipage{0.2\textwidth}
        \includegraphics[width=\linewidth]{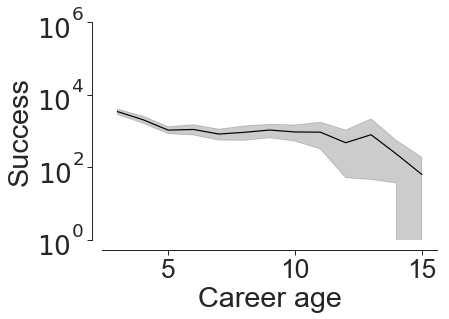}
        \subcaption{Stable unsuccessful}\label{fig2:career_cat2}
    \endminipage\hfill
    \minipage{0.2\textwidth}%
        \includegraphics[width=\linewidth]{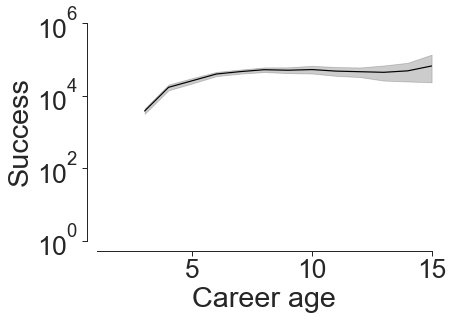}
        \subcaption{Upward}\label{fig:career_upwards}
    \endminipage\hfill
    \minipage{0.2\textwidth}%
        \includegraphics[width=\linewidth]{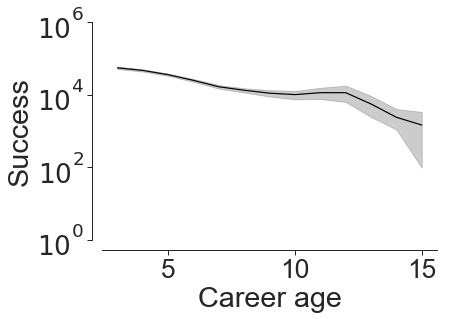}
        \subcaption{Downward}\label{fig:career_downward}
    \endminipage
    \caption{Artists can be grouped into five categories according to their career trajectories. Curves depict the average travel distance with 95\% confidence interval. The number of artists within each group are (left to right): $1362$, $394$, $985$, $393$, and $1090$.}
    \label{fig:career_types}
\end{figure*}
We report the marginal (just fixed effects) and conditional (fixed and random effects) pseudo-coefficients of determination ($R^2$) \citep{nakagawa2013general}, the Akaike Information Criterion (AIC), and several statistics related to the random effects.
We only keep observations that contain both release and live performance activities.
These filters reduce the number in the overall data set (table \ref{tab:data_stats}) to 4,224 musicians and 27,077 observations (musician-career age combinations).

We use linear mixed models \citep{bates2014fitting} with musicians and cohorts as random effects. Musicians differ in individual characteristics like skills or creativity. In addition, they are likely effected by cohort-specific conditions. For example, musicians whose start year coincides with the widespread popularity and internationalization of EDM culture are likely to have a higher average number of gigs and longer travel trajectories. Similarly, self-promotion on digital platforms is a rather new practice.
By fitting musicians and start years as random effects, we account for variations within these variables.
Independent variables are $z$-standardized, i.e., they indicate how much a score of socio-cultural capital deviates (in terms of standard deviations) from the average value of all musicians in a year $t$. This transformation partially accounts for the year-specific variations in the field.
The dependent variable is logged.
Career analysis is implemented via interaction effects. We use a musician's career stage as an interaction term with dummy coding to evaluate the association of each independent variable with success at different phases of a career.

\section*{Results}

\begin{table*}[t!]
\centering
\begin{minipage}[t]{0.45\linewidth}
\begin{tabular}[t]{lll}

\toprule
 & \bf Model~1 & \bf Model~2 \\
\midrule
Intercept         & \bf 9.007   & \bf 8.971   \\
Number of gigs       & \bf .454  & .049       \\
Number of releases       & \bf .132  & \bf .097 \\
Mid career    & \bf -1.355 & \bf -1.324\\
Late career       & \bf -2.090  & \bf -2.233 \\
Number of gigs*mid career                &                                & \bf .499 \\
Number of gigs*late career               &                                & \bf .608   \\
Number of releases*mid career     &    & .045       \\
Number of releases*late career   &  & .217       \\
\midrule
\bf Co-gig &  &  \\
~~Community    & \bf 1.210 & \bf 1.188  \\
~~Mainstream     & \bf .328   & \bf .514    \\
~~Bridging      & -.046    & -.042     \\
~~Constraint     & \bf -.288   & \bf -.476\\
~~Community*mid career      &  &.117 \\
~~Community *late career    &      & \bf .560$\asti$  \\
~~Mainstream*mid career     &  & \bf -.321  \\
~~Mainstream*late career                   &                                & \bf -.627   \\
~~Bridging*mid career       &   & -.003     \\
~~Bridging *late career        &         & -.024     \\
~~Constraint*mid career                     &                                & \bf .379   \\
~~Constraint*late career                    &                                & .171    \\
\midrule
\bf Co-venue &  &  \\
~~Community   & -.005     & -.084    \\
~~Mainstream  & .071  & -.020      \\
~~Bridging  & .003      & .002     \\
~~Constraint    & .008    & \bf .104    \\
~~Community*mid career     &  &  \bf .125   \\
~~Community*late career   &    & .247  \\
~~Mainstream*mid career     &    & \bf .168  \\
~~Mainstream*late career                    &                                & \bf .373   \\
~~Bridging*mid career                     &                                & .005   \\
~~Bridging*late career      &    & -.06   \\
~~Constraint*mid career     &   & \bf -.172 \\
~~Constraint*late career      &    & \bf -.245 \\
\midrule
\end{tabular}%
\end{minipage} 
\quad
\begin{minipage}[t]{0.45\linewidth}
\begin{tabular}[t]{lll}
\toprule
 & \bf Model 1 & \bf Model 2 \\
\midrule
\bf Co-label &  &  \\
~~Community    & .023       & .037      \\
~~Mainstream    & .045 & \bf .095   \\
~~Bridging    & .009 & .027       \\
~~Constraint   & -.016  & -.013    \\
~~Community*mid career    &     & -.022      \\
~~Community*late career        & & -.058     \\
~~Mainstream*mid career         &  & -.080     \\
~~Mainstream*late career   &      & -.083     \\
~~Bridging*mid career      &  & -.040      \\
~~Bridging*late career    &    & .067      \\
~~Constraint*mid career    &        & .000       \\
~~Constraint*late career        &     & -.082  \\
\midrule
\bf Co-style &  &  \\
~~Community  & \bf -.117 $\astii$   & \bf -.099 $\astii$   \\
~~Mainstream   & \bf .092 $\astii$    & .106     \\
~~Bridging   & .007      & -.002      \\
~~Constraint    & .001   & -.057   \\
~~Community*mid career       &    & -.012   \\
~~Community*late career     &  & \bf -.277 $\astii$  \\
~~Mainstream*mid career      &      & -.010     \\
~~Mainstream*late career   &     & -.192    \\
~~Bridging*mid career       &  & .036      \\
~~Bridging*late career      &      & -.136    \\
~~Constraint*mid career       &   & .100    \\
~~Constraint*late career      &       & .088     \\
\midrule
Marginal R$^2$                   & .242 & .245  \\
Conditional R$^2$                & .617  & .625 \\
AIC                              & 137,811   & 137,613  \\
Variance: Musicians (Intercept)          & 6.753 & 6.875\\
Variance: Start years (Intercept) & .097 & .100  \\
Variance: Residual       & 6.989  & 6.891  \\
\bottomrule
\end{tabular}
\end{minipage}
\caption{Results of mixed-effects regressions of success. Model 1 contains baseline and network-based variables, model 2 also includes interactions with career stage dummy variables. Independent variables and their interactions computed for moving 3-year time windows explain success in the ensuing 3 years. Effect sizes are log odds ratios (i.e., for a one-unit increase in an independent variable $x$, there is a $\exp(x)-1$ percent increase in the likelihood of success). In model 2, variables without an interaction term represent the population average effect. For the interpretation of interaction effects, coefficients must be summed (example in the text). Intervals are reported for the 95\% confidence level. Bold coefficients indicates that Null hypothesis value is outside the confidence interval. \textbf{$\asti$} indicates effect is not significant in corresponding model that excludes release-based variables. \textbf{$\astii$} indicates Effect is not significant in corresponding model that excludes performance-based variables.} \label{tab:results}
\end{table*}

Table \ref{tab:results} reports the results from two regression models, where interaction terms for career stages are added in the second one. To ease understanding, we report the percent changes that can be obtained from the table. First, we report the results regarding non-network variables. Then, answers to the four research questions are given in dedicated subsections whose headlines sum up the main answers.

The baseline model 1 shows that a one-standard-deviation increase in the number of gigs increases the likelihood of success by a factor of $\exp(0.454)=1.57$ (a 57\% increase). Releasing more music, on the other hand, is less associated with increased success (14\%).
The largest effects we find pertain to how success changes as musicians advance in their careers. Musicians in the mid career stage are $73\%$ less likely and musicians in the late career stage are even $88\%$ less likely to be successful than early-career musicians. That means, success is mostly an early-career phenomenon.
One explanation is that the dependent variable is a travel-based proxy of success. The finding then is that musicians travel less the more their career advances. However, figure \ref{fig:career_types} shows that decreases of success with career age are just the average effect. In fact, there are quite a few musicians with stable successful and even upward career trajectories.
Correspondingly, when we consider interaction effects (model 2), we find that playing more gigs is associated with larger increases of success the more careers advance:
Although the impact of number of gigs in early career is not clear, it is likely to increase the chance of success in mid and late career dramatically. With this in mind, we move on to answering the research questions.
The first one asks about the association of success with community embeddedness.

\subsection*{Successful Musicians Embed Into Communities at Gigs}

Embedding into communities that result from social relations at gigs is most strongly associated with success (235\% increase, model 1), particularly in the late career stage (474\% increase, model 2).
Co-venue networks are indicative of the importance of place. Due to our bipartite network approach, musicians that perform in core venues are core musicians in co-venue networks. We find that performing in core venues becomes significantly more important in the mid-career stage, but the effect is very small (4\% increase).
There are also significant effects regarding the importance of music style. Interestingly, community embeddedness is negatively associated with success (11\% decrease, model 1), with decreases rising from 9\% in the early career to 31\% in the late career stage (model 2).

The conditional $R^2$ states that model 2 can explain 62.5\% of the variance in success. But since the marginal $R^2$ is at 24.5\%, most of the variance is explained by individual characteristics which we do not measure. Also, the marginal $R^2$ of models (performed as robustness checks, not reported here) that exclude performance-based variables (number of gigs, co-gig and co-venue network variables) is a mere 2.8\%. That means, the practice of releasing music is practically not relevant for success, while most explanatory power comes from live performances. Correspondingly, no effects related to music styles are robust.

To answer the first research question, we found that only communities formed at gigs are associated with success, but strongly so. 
We next contextualize this result by answering how success is associated with mainstream belonging. Is success all about alternative communities, as the literature suggests? Or is embedding into mainstream communities a path to success after all?
\subsection*{Successful Musicians Avoid Mainstream Gigs but Seek Mainstream Venues Over Time}

Belonging to the co-gig mainstream is associated with an overall 38\% increase in success (model 1), but there is a significant negative trend over an average career. The effect is strongest in the early career (67\% increase, model 2) but becomes modest in the mid career (21\% increase) and even turns into a 11\% decrease in the late career stage.
Opposing this trend, performing in mainstream venues is slightly associated with success increases after the early career stage. They amount to 16\% and 42\% increases in the mid and late career stages, respectively.
There are very small effects ($<10\%$ increases of success) that releasing on mainstream labels and in mainstream styles is beneficial in the early career stage. However, for lack of explanatory power and robustness we will not discuss these.

The answer to the second research question, thus, is that mainstream belonging actually is associated with success with opposing trends for gig-based and venue-based networks.
The emerging picture is that, while embedding into gig-based communities is important throughout successful careers, these communities transform from mainstream to alternative communities as careers progress (or musicians move between them accordingly).
Whereas the first two questions detailed the role of network closure, we next investigate network openness. The third research question asks about the association of success with bridging.

\subsection*{Successful Musicians are at Home in One Exclusive Community}

Bridging otherwise disconnected parts is never associated with success in any of the four networks. From the perspective of the general networks literature, this is surprising because bridging positions are often found to be sources of creativity. However, from the perspective of the EDM literature, this null result is perfectly expected. It indirectly suggests that the communities that successful musicians embed into have an exclusive character. In other words, positions in multiple, or between, communities are not rewarded. Successful musicians are at home in one community that is walled off from others.
This finding begs the question of whether musicians need to distinguish themselves while belonging to one, exclusive community to find individual success.

\subsection*{Successful Musicians Avoid Redundant Connections at Gigs}

As expected, dense co-gig ego networks have a constraining effect (25\% decrease of success). Adding interaction effects does not yield a trend over career stages (decreases jump from 38\% to 9\% and 26\%).
Turning to venue-based network variables, high constraint means that musicians cluster by playing in a redundant and, hence, indistinguishable set of venues in terms of musicians playing there. Constraint turns from making success slightly likely in the early career (11\% increase) to making it slightly unlikely in the mid career (7\% decrease) and late career (13\% decrease) stages.

The answer to the last research question is that too dense ego networks constrain success in all performance-based networks and career stages. The exception is that it is beneficial to start careers by playing in venues that host a redundant set of musicians.

\section*{Discussion}

\subsubsection*{Summary}

Field theory posits that agents in markets strive for revenue while agents in artistic fields strive for autonomy. However, many artists in EDM do strive for commercial success \citep{reitsamer2011diy}. This hipster paradox creates a dilemma. On the one hand, artists strive to make a living from performing live and releasing music; on the other hand, commercial success is collectively despised due to the counter-cultural roots of EDM. It has been proposed that musicians solve this dilemma by embedding into alternative and exclusive communities in which work and play fuses \citep{wreo20632}.

We find that embedding into communities that derive from social relations at live gigs is, indeed, most strongly associated with success for an average musician. This is particularly the case in the late career stage where, on average, success tends to decrease.
However, in the early and mid career stages, it is mainstream communities in the core of the field, not alternative communities in the periphery, that increase the likelihood of success. It is only in the late career stage that mainstream belonging is negatively associated with success. This finding gives nuance to the explanation that the autonomous way of embedding into alternative communities is the path to success all the way through.
Yet, we do find indirect empirical evidence that distancing from others is important as positions between communities are never associated with success. Boundaries around exclusive communities, in other words, matter.
In addition to all explanations proposed so far, we find that it is also important that gig communities avoid redundancy so that musicians can leverage the creative potential of varied contacts.

Our findings become even more nuanced if we contrast gigs with venues. Venues are known to be drivers of communities where musicians meet their exclusive crowds \citep{lange_value_2013}. Here, we find weak evidence for a crossover effect. In the early career stage, successful musicians play in venues that host a redundant community of artists. As their careers progress, it becomes increasingly important to perform in the mainstream venues of the field.
Finally, by releasing music artists demonstrate their seriousness \citep{mcleod2001genres}. We do not find this practice to contribute to an explanation of travel-based success.

In sum, our results constitute evidence of a structural trade-off among revenue and autonomy. Musicians in EDM embed into exclusive performance-based communities for autonomy but, in earlier career stages, seek the mainstream for commercial success.

\subsubsection*{Limitations and Future Work}

Our dependent success variable is derived from travel trajectories and is, hence, a proxy for success. We have done so because, in our research design, success must be measured for rolling time windows.
In principle, success can also be defined in various other ways such as record sales, record label deals, prices, or online popularity such as the number of followers on social media platforms. If historic data can be leveraged, future studies could use different non-proxy metrics or combine multiple metrics in a compound measure. 

This study considers the practices of performing and releasing. However, self-promotion is becoming ever more important \citep{allington2015networks, wreo20632}. Social media platforms such as \textit{SoundCloud} and \textit{Instagram} allow artists not only to promote themselves on a global scale, but also to connect and interact with their peers in new ways.
What is more, our study is restricted to the production side of cultural objects. But their consumption also leaves digital traces, for example, in the form of likes, mentions, and purchases. 
Future studies could also account for the impact of self-promotion and cultural consumption.

The dataset comes with a number of limitations. For the most, the recency and self-selection in RA may bias the results of this study to certain musicians and music genres and a particular time period. For example, the number of gigs, artists and venues that register in the website show an exponential increase over time. 
The self-selection results in over representation of certain artists. For example Tech house, Techno, Minimal, Deephouse, and House account for more than 50\% of releases and events in the datasets. 
Furthermore, name ambiguity, inaccurate and faulty content, and APIs could introduce errors in our dataset. However, our manual evaluations show these errors are marginal.

\section*{Conclusion}

We have studied the hipster paradox as it yields an interesting behavioral dilemma in the field of EDM.
Our results support the explanation offered by the EDM literature, namely, that musicians embed into exclusive performance-based communities to be autonomous in their quest for success.
Our longitudinal study allows to refine this explanation since we find behavioral differences between musicians in different career stages. In earlier career stages, musicians seek the mainstream for commercial success.
Cultural production in the field of EDM cannot be explained by a polar distinction between art and money. Instead, our results point towards a structural trade-off among revenue and autonomy.

We hope our approach highlights that large-scale digital behavioral data, together with computational methods and social theories, allow to gain new insights into social phenomena such as the hipster paradox. Besides the explanations offered in the EDM literature, we also relied on general theories like field and network theory and the formal methods they provide. We believe that our approach is quite generic and can be used to study other fields of cultural productions, especially music.

\section*{Acknowledgements}

We would like to thank Andreas Schmitz and Maria Zens for inspiring conversations and their valuable inputs on conceptual and methodological consideration of field theory.

\bibliography{ref.bib}

\end{document}